\documentclass[twocolumn]{aastex631}
\graphicspath{{./fig/}{./png/}}
\usepackage{bm}

\newcommand{\EQ}{\begin{equation}}
\newcommand{\EN}{\end{equation}}
\newcommand{\EQA}{\begin{eqnarray}}
\newcommand{\ENA}{\end{eqnarray}}

\newcommand{\Eq}[1]{Equation~(\ref{#1})}
\newcommand{\Eqs}[2]{Equations~(\ref{#1}) and~(\ref{#2})}

\newcommand{\App}[1]{Appendix~\ref{#1}}

\newcommand{\Fig}[1]{Figure~\ref{#1}}

\newcommand{\Tab}[1]{Table~\ref{#1}}

\newcommand{\bbra}[1]{\left\langle #1\right\rangle}

\newcommand{\meanrho}{\overline{\rho}}

{}
{}
\newcommand{\meanFFFF}{\overline{\mbox{\boldmath ${\cal F}$}}{}}{}

\newcommand{\meanemf}{\overline{\cal E} {}}

{}
{}
\newcommand{\meanEMF}{\overline{\mbox{\boldmath ${\cal E}$}}{}}{}
{}
{}
{}
{}
{}
{}
{}
{}
{}
{}
{}
{}
{}
{}
\newcommand{\meanUU}{\overline{\bm{U}}}

{}

\newcommand{\meanB}{\overline{B}}

\newcommand{\meanC}{\overline{C}}

\newcommand{\meanS}{\overline{S}}
\newcommand{\meanT}{\overline{T}}

\newcommand{\meanFFF}{\overline{\cal F}}

\newcommand{\hatkk}{\hat{\bm{k}}}

{}

{}
{}
{}

\newcommand{\meanBB}{{\overline{\bm{B}}}}

\newcommand{\kk}{\bm{k}}

\newcommand{\bb}{\bm{b}}
\newcommand{\BB}{\bm{B}}

\newcommand{\EE}{\bm{E}}

\newcommand{\UU}{\bm{U}}
\newcommand{\FF}{\bm{F}}

\newcommand{\uu}{\bm{u}}

\newcommand{\ee}{\mbox{\boldmath $e$} {}}

\newcommand{\ff}{\mbox{\boldmath $f$} {}}

\newcommand{\FFF}{\mbox{\boldmath ${\cal F}$} {}}

\newcommand{\nab}{{\bm{\nabla}}}

\newcommand{\ggamma}{\mbox{\boldmath $\gamma$} {}}

\newcommand{\SSSS}{\mbox{\boldmath ${\sf S}$} {}}

\newcommand{\EMF}{\mbox{\boldmath ${\cal E}$} {}}

\newcommand{\DD}{{\rm D} {}}

\newcommand{\dd}{{\rm d} {}}

\def\ga{\mathrel{\mathchoice {\vcenter{\offinterlineskip\halign{\hfil
$\displaystyle##$\hfil\cr>\cr\sim\cr}}}
{\vcenter{\offinterlineskip\halign{\hfil$\textstyle##$\hfil\cr>\cr\sim\cr}}}
{\vcenter{\offinterlineskip\halign{\hfil$\scriptstyle##$\hfil\cr>\cr\sim\cr}}}
{\vcenter{\offinterlineskip\halign{\hfil$\scriptscriptstyle##$\hfil\cr>\cr\sim\cr}}}}}
\def\H{\mbox{\rm H}}

\def\Ma{\mbox{\rm Ma}}
\def\Co{\mbox{\rm Co}}

\def\Sc{\mbox{\rm Sc}}

\def\Pm{\mbox{\rm Pr}_{\rm M}}
\def\Rm{\mbox{\rm Re}_{\rm M}}

\def\Rey{\mbox{\rm Re}}

\def\Pe{\mbox{\rm Pe}}
\def\Co{\mbox{\rm Co}}
\def\Gr{\mbox{\rm Gr}}

\def\tauc{\tau_{\rm c}}

\def\EK{E_{\rm K}}

\def\cp{c_{\rm p}}
\def\cv{c_{\rm v}}

\def\cs{c_{\rm s}}

\def\kf{k_{\rm f}}

\def\epsf{\epsilon_{\rm f}}

\def\epsK{\epsilon_{\rm K}}

\def\urms{u_{\rm rms}}

\def\kappat{\kappa_{\rm t}}

\def\chit{\chi_{\rm t}}

\def\etat{\eta_{\rm t}}

\newcommand{\W}{\,{\rm W}}
\newcommand{\V}{\,{\rm V}}

\newcommand{\T}{\,{\rm T}}
\newcommand{\G}{\,{\rm G}}

\newcommand{\K}{\,{\rm K}}
\newcommand{\g}{\,{\rm g}}
\newcommand{\s}{\,{\rm s}}

\newcommand{\m}{\,{\rm m}}

\newcommand{\J}{\,{\rm J}}

\newcommand{\A}{\,{\rm A}}

\def\apjl{ApJL}
\def\apjs{ApJS}

\def\aap{A\&A}

\begin{document}

\title{Helicity effect on turbulent passive and active scalar diffusivities}
\email{brandenb@nordita.org}
\author[0000-0002-7304-021X]{Axel Brandenburg}
\affiliation{Nordita, KTH Royal Institute of Technology and Stockholm University, Hannes Alfv\'ens v\"ag 12, SE-10691 Stockholm, Sweden}
\affiliation{The Oskar Klein Centre, Department of Astronomy, Stockholm University, AlbaNova, SE-10691 Stockholm, Sweden}
\affiliation{McWilliams Center for Cosmology \& Department of Physics, Carnegie Mellon University, Pittsburgh, PA 15213, USA}
\affiliation{School of Natural Sciences and Medicine, Ilia State University, 3-5 Cholokashvili Avenue, 0194 Tbilisi, Georgia}

\author[0000-0001-9619-0053]{Petri J. K\"apyl\"a}
\affiliation{Institut f\"ur Sonnenphysik (KIS), Georges-K\"ohler-Allee 401a, 79110 Freiburg im Breisgau, Germany}

\author[0000-0001-7308-4768]{Igor Rogachevskii}
\affiliation{Department of Mechanical Engineering, Ben-Gurion University of the Negev, P.O. Box 653, Beer-Sheva 84105, Israel}
\affiliation{Nordita, KTH Royal Institute of Technology and Stockholm University, Hannes Alfv\'ens v\"ag 12, SE-10691 Stockholm, Sweden}

\author[0000-0002-5242-7634]{Nobumitsu Yokoi}
\affiliation{Institute of Industrial Science, University of Tokyo, Komaba, Meguro, Tokyo 153-8505, Japan}

\begin{abstract}
Turbulent flows are known to produce enhanced effective magnetic and
passive scalar diffusivities, which can fairly accurately be determined
with numerical methods.
It is now known that, if the flow is also helical, the 
effective magnetic diffusivity is reduced
relative to the nonhelical value.
Neither the usual second-order correlation approximation
nor the various $\tau$ approaches
have been able to capture this.
Here we show that the helicity effect on the turbulent passive scalar diffusivity
works in the opposite sense and leads to an enhancement.
We have also demonstrated that the correlation time of the turbulent velocity field increases with the kinetic helicity.
This is a key point in the theoretical interpretation
of the obtained numerical results.
Simulations in which helicity is being produced self-consistently by stratified rotating turbulence resulted in a turbulent passive scalar diffusivity
that was found to be decreasing with increasing rotation rate.
\end{abstract}

\section{Introduction}

In many astrophysical plasmas such as stellar convection zones, the
interstellar medium, and accretion discs, the Reynolds numbers are
extremely large.
Therefore, to describe the large-scale behavior of such flows, one often
replaces the small viscosity or diffusion coefficients by effective ones.
Turbulent diffusivities in the evolution equations for passive scalars
act similarly as ordinary (molecular or atomic) ones, except that they
characterize the diffusion of larger scale structures, as described by
the corresponding averaged or coarse-grained evolution equations.
Denoting the mean passive scalar concentration $C$ by an overbar, the
equation for $\meanC$ is given by
\begin{equation}
\frac{\partial\meanC}{\partial t}=-\nab\bm\cdot\left(\meanUU\,\meanC\right)
+(\kappa+\kappat)\nabla^2\meanC,
\label{Eq1}
\end{equation}
where we have allowed for the possibility of a mean flow $\meanUU$, while
$\kappa$ and $\kappat$ are the microphysical and turbulent diffusion
coefficients, respectively.
The diffusion coefficients are proportional to the product of the mean-free
path and the typical velocity of particles or, in the turbulent case, 
the product of the integral turbulent scale and the rms velocity.
Equation~(\ref{Eq1}) is written for turbulence without stratification of the mean density or temperature so that
effective pumping velocity caused by the turbulent thermal diffusion vanishes \citep{EKR97,RI21}.

The derivation of the turbulent diffusion coefficients is usually done by some approximations.
Meanwhile, significant progress has been made by numerically computing
these turbulent coefficients.
A particularly useful approach is the test-field method \citep{Sch05,
Sch07}, which was originally applied to magnetic fields in spherical
geometry and then to Cartesian domains \citep{Bra05,BRS08}.
This method is sufficiently accurate to identify subtle effects caused
by kinetic helicity in the flow \citep{BSR17}.

In the presence of magnetic fields, the kinetic helicity
causes completely new qualities of its own.
Unlike the case of turbulent or microphysical diffusion, helicity also produces nondiffusive effects that lead to a destabilization of the nonmagnetic state.
This is because helicity is a pseudoscalar, which can couple the axial magnetic field vector with the polar electric field vector to give an
extra contribution to the turbulent electromotive force in the mean-field induction equation.
By contrast, the turbulent magnetic diffusivity is an ordinary scalar.
It was therefore surprising when kinetic helicity was found to affect even the turbulent magnetic diffusivity \citep{BSR17}.
This effect was such that helicity suppresses the turbulent magnetic diffusivity by a certain amount.
The possibility of a helicity effect on the turbulent magnetic diffusivity
was already noticed in the early work of \cite{NS88}, but 
they found an enhancement of the turbulent magnetic diffusivity by the kinetic helicity.

Applying the Feynman diagram technique, \cite{DS87} show that kinetic helicity can increase the turbulent diffusion of a passive scalar field.
On the other hand, subsequent work by \cite{Zhou90} using renormalization-group theory found no effect of helicity on the renormalized eddy viscosity.
The effect of kinetic helicity on passive scalar diffusion was also investigated by \cite{Chkhetiani+06} using the renormalization-group approach.
They found that the effective diffusivity can be 50\% larger in the helical case.
They also noted that there is no helicity effect on the anomalous scaling of the structure functions.

The results of \cite{BSR17} were recently verified by \cite{Mizerski23} 
using the renormalization-group approach.
In particular, he found that for small magnetic Reynolds numbers, the helical correction to turbulent diffusion of 
the mean magnetic field is proportional to
$\Rm^2 (H_{\rm K} \tau_{\rm c})^2 / \langle {\bm u}^2 \rangle$, where $\Rm = \tau_{\rm c} \, \langle {\bm u}^2 \rangle/\eta$ is the magnetic Reynolds number,
$\tau_{\rm c}$ is the turbulent correlation time, $\eta$ is the magnetic diffusion caused by an electrical conductivity of plasma,
and $H_{\rm K} =\langle{\bm u} \cdot  {\bm \omega}  \rangle$ is the kinetic helicity.
This scaling $(\propto \Rm^2)$ is shown in Figure~4 of \cite{BSR17}.
This confirms that the helical correction cannot emerge from the second-order correlation approximation, 
where the transport coefficients are only linear in the magnetic Reynolds number.

What has not yet been specifically addressed is the effect of helicity on the passive scalar diffusivity or even the thermal diffusivity of an active scalar such as the temperature or the specific entropy in the
mean-field energy equation.
Doing this is the purpose of the present work.

Helicity affects the value of the turbulent passive and active scalar diffusivity in a clear and consistent way.
This is similar to the helicity effect on the turbulent magnetic diffusivity, but this new effect is the other way around, i.e., the turbulent passive and active scalar diffusion are enhanced by helicity, while the turbulent magnetic diffusivity is decreased.
In the accompanying theoretical paper
by \cite{RKB25}, remaining puzzles are addressed and possible explanations are being proposed.

Of some interest in this context is the earlier work of \cite{BRK12}, who computed turbulent magnetic field and passive scalar transport for rotating stratified turbulence.
The combined presence of rotation and stratification also leads to helicity and therefore to an $\alpha$ effect.
They found a slight decrease of the magnetic diffusivity as the angular velocity is increased.
At the time, this was not thought to be surprising because the focus was on new turbulent transport coefficients that only arise because of rotation and stratification.
Furthermore, already rotation alone (without helicity) is known to decrease the turbulent magnetic diffusivity \citep{RKR03}.

For most astrophysical purposes, only order-of-magnitude estimates of
turbulent transport coefficients are usually considered.
This may change in future, when more accurate methods and measurements become more commonly available both in simulations and in observations.
For example, the discrepancy in the estimate for the turbulent magnetic diffusivity was noticed in theoretical work in high-energy astrophysics on the chiral magnetic effect when simple estimates for the turbulent magnetic diffusivity did not match previous estimates \citep{Schober+18}.
This discrepancy was then explained by the presence of helicity in one of the cases.

\section{Our model}

We consider both isothermal and nonisothermal turbulence and begin with the former.

\subsection{Basic equations for isothermal turbulence}

Our basic equations are the induction and passive scalar equations for
the magnetic field $\BB$ and the passive scalar 
concentration $C$ (e.g., number density of particles).
The magnetic field is also divergence free.
The governing equations are then
\begin{equation}
\frac{\partial\BB}{\partial t}=\nab\times\left(\UU\times\BB-\EE_\mathrm{diff}\right),
\quad \EE_\mathrm{diff}=-\eta\nab\times\BB,
\end{equation}
\begin{equation}
  \frac{\partial C}{\partial t}=\nab\bm\cdot\left(-\UU C-\FF_\mathrm{diff}\right),
\quad \FF_\mathrm{diff}=-\kappa\nab C.
\end{equation}

The velocity $\UU$ is obtained as a solution of the Navier-Stokes equations.
In the kinematic test-field method, we ignore the feedback of the magnetic
field on the flow, i.e., we solve
\begin{equation}
\frac{\DD\UU}{\DD t}=-\cs^2\nab\ln\rho+\ff+\frac{1}{\rho}\nab\bm\cdot(2\rho\nu\SSSS),
\end{equation}
\begin{equation}
\frac{\DD\ln\rho}{\DD t}=-\nab\bm\cdot\UU.
\end{equation}
where $\rho$ is the density, $\cs$ is the isothermal sound speed,
$\nu$ is the kinematic viscosity, $\SSSS$ is the rate of strain tensor
with the components ${\sf S}_{ij}=(\partial_i u_j+\partial_j u_i)/2
-\delta_{ij}\nab\bm\cdot\uu/3$, and $\ff$ represents a forcing function
that is $\delta$ correlated in time and consists of plane waves with a
mean forcing wavenumber $\kf$.
It is given by $f_i=R_{ij}f_j^\mathrm{(nohel)}$, where
$R(\hatkk)=(\delta_{ij}-\sigma\epsilon_{ijk}\hat{k}_k)/\sqrt{1+\sigma^2}$
depends on $\hatkk=\kk/k$ with $k=|\kk|$ and the fractional helicity
$\sigma$, and $\ff^\mathrm{(nohel)}=f_0\,\ee\times\kk/|\ee\times\kk|$
is a nonhelical forcing function with $f_0$ being a scaling factor and
$\ee$ a random vector that is not aligned with $\kk$.

\subsection{Equations for nonisothermal turbulence}

In our simulations of nonisothermal turbulence, we measure the response
of the system to imposing large-scale gradient of specific
entropy $s$ with a relaxation time $\tau$.
The evolution equations for $\uu$ and $s$ are then
\begin{equation}
\frac{\DD\UU}{\DD t}=-\cs^2\nab(\ln\rho+s/\cp)+\ff+\frac{1}{\rho}\nab\bm\cdot(2\rho\nu\SSSS),
\end{equation}
\begin{equation}
T\frac{\DD s}{\DD t}=2\nu\SSSS^2+\frac{1}{\rho}\left(\nab\bm\cdot\FF_\mathrm{rad}-\mathcal{C}\right)-\frac{s-\tilde{s}_0}{\tau},
\end{equation}
where $T$ is the temperature, $\cp$ is the specific heat at constant
pressure, $\FF_\mathrm{rad}=-\cp\rho\chi\nab T$ is the radiative flux,
and $\mathcal{C}$ is a volumetric cooling function to compensate for
viscous heating.
Since the system is no longer isothermal, the sound speed is now given by
$\cs^2=(\gamma-1)\cp T$, where $\gamma=\cp/\cv$ is the ratio of specific
heats and $\cv$ is the specific heat at constant volume.
For the target profile of specific entropy, we choose
$\tilde{s}_0=s_0\sin k_T z$.
Here, we take $k_T=k_1$ for what will later be called the test-field wavenumber, where $k_1=2\pi/L$ is the smallest wavenumber in the domain.
Different values of $k_T$ would be of interest for studying the scale dependence of turbulent transport, as has been done 
on various occasions \citep{BS02, BRS08, BSV09}.

\subsection{Parameters}

For the scale separation ratio, i.e., the ratio of the forcing wavenumber $\kf$ and the box wavenumber $k_1$, we take $\kf/k_1=5.1$ in most of our cases.
Although not stated explicitly there, this was also the value adopted in \cite{BSR17}.
Larger (smaller) values of $\kf$ allow us to access larger (smaller) scale separation ratios.
At the end of this paper, we present a small survey of different choices; see also \cite{BRS08, BSV09} for such studies in other contexts.
Our main governing control parameters are the Reynolds number $\Rey=\urms/\nu\kf$ and the Mach number $\Ma=\urms/\cs$.
The Schmidt number, $\Sc=\nu/\kappa$, the magnetic Prandtl number, $\Pm=\nu/\eta$, and the thermal Prandtl number $\Pr=\nu/\chi$ are unity in all cases.
Therefore, the magnetic Reynolds number $\Rm=\urms/\eta\kf$ and the P\'eclet number $\Pe=\urms/\chi\kf$ equal the fluid Reynolds number in all cases.

\subsection{Test-field methods}

The test-field method implies the simultaneous solution of additional
equations for the fluctuating magnetic field or the fluctuating passive
scalar concentration.
The variables are indicated by the letter $T$.
The equations are obtained by subtracting the corresponding averaged equations
from the original ones and yield
\begin{equation}
\frac{\partial\bb^T}{\partial t}=\nab\times\left(
\uu\times\meanBB^T+\meanUU\times\bb^T+\EMF_T'\right)
+\eta\nabla^2\bb^T,
\label{dbTdt}
\end{equation}
\begin{equation}
  \frac{\partial c^T}{\partial t}=\nab\bm\cdot\left(
-\uu\meanC^T-\meanUU c^T+\FFF_T'\right)
+\kappa\nabla^2 c^T,
\label{dcTdt}
\end{equation}
where $\EMF_T'=\uu\times\bb-\overline{\uu\times\bb}$ and
$\FFF_T'=-(\uu c-\overline{\uu c})$ are nonlinear terms that are neglected
in the second-order correlation approximation.
Including those terms yields the new subtle effects that we found in
\cite{BSR17} for $\etat$ and in the present work for $\kappat$.

In the following, we assume planar averages and denote them by overbars,
e.g., $\meanBB(z,t)=\int\BB(x,y,z,t)\,\dd x\,\dd y/L_\perp^2$, where $L_\perp$
is the extent of the computational domain in the $xy$ plane.
In the spirit of the test-field method, one decouples \Eqs{dbTdt}{dcTdt}
from those for the actual fluctuations and solve them for a set of mean
fields (mean scalars) such that one can compute $\alpha_{ij}$, $\eta_{ij}$,
and $\kappa_{ij}$ uniquely for each time step and at each value of $z$.
Using as a shorthand $s=\sin k_Tz$ and $c=\cos k_Tz$,
we choose sinusoidal and cosinusoidal test fields $\meanBB^1=(s,0,0)$,
$\meanBB^2=(c,0,0)$, $\meanBB^3=(0,s,0)$, and $\meanBB^4=(0,c,0)$, as
well as $\meanC^1=s$ and $\meanC^2=c$, i.e., four different test fields
for $\meanBB^T$ and two different ones for $\meanC^T$.
This allows us to compute the coefficient $\alpha_{ij}$, $\eta_{ij}$,
and $\kappa_{ij}$ in the parameterizations
\begin{equation}
\meanemf_i^T=\alpha_{ij}\meanB_j^T-\eta_{ij}(\nab\times\meanBB^T)_j,
\end{equation}
\begin{equation}
\meanFFF_i^T=\gamma_i\meanC^T-\kappa_{ij}\nabla_j\meanC^T,
\end{equation}
where $\meanEMF^T=\overline{\uu\times\bb_T}$, $\meanFFFF^T=-\overline{\uu c_T}$,
and $i,j=1,2$ denote the $x$- and $y$-components.
The aforementioned turbulent viscosity and passive scalar diffusivity are then
given by $\etat=(\eta_{11}+\eta_{22})/2$ and $\kappat=(\kappa_{11}+\kappa_{22})/2$.
The effective pumping velocity $\ggamma$ 
of the mean magnetic field vanishes for
homogeneous turbulence,
but the effective pumping velocity $\ggamma$ 
of the mean passive scalar field due to the density stratification of the fluid \citep{EKR97,RI21}
was found to lead to downward transport of the mean passive scalar concentration (to the maximum of the mean fluid density)
in density-stratified turbulence \citep{BRK12,HKRB12}.

It should be noted that in the original application of the test-field method, \cite{Sch05, Sch07} used a combination of constant and linearly varying test fields.
This choice is appropriate for computing turbulent transport properties on the largest possible scales, but it is not well suited for the use in periodic domains.
This was the main reason why \cite{Bra05} employed sinusoidal and cosinusoidal test fields,
but it also provided a natural way of computing the dependence of the turbulent transport coefficients on different length scales or for different wavenumbers.
The resulting formulation of the electromotive force in Fourier space translates directly into one in terms of integral kernels \citep{BRS08}.
This allowed us to avoid the restriction to large scale separation in space and time by replacing the multiplications with turbulent transport coefficients
by a convolution with the appropriate integral kernels; see \cite{HB09} and \cite{RB12} for corresponding studies.
The effect of different spatial scales on turbulent mixing was also investigated by \cite{deAvillez+MacLow02} using checkerboard patterns,
but this approach cannot so easily be utilized in the framework of mean-field theory.

\subsection{Active scalar diffusivity}

To determine the turbulent radiative diffusion coefficient, we use the
standard mean-field expression for the enthalpy flux \citep{Ruediger89},
\begin{equation}
\meanFFFF_\mathrm{enth}=-\chit\meanrho\meanT\nab\meanS,\label{equ:Fenth}
\end{equation}
where the actual enthalpy flux is computed as
$\meanFFFF_\mathrm{enth}=\overline{(\rho\UU)'\cp T'}$, and correlate
their $z$-components against each other to determine $\chit$.
Here, primes denote the departures from the horizontal means.
This method follows that employed by \cite{KS22}, who also computed the
turbulent kinematic viscosity in an analogous way be correlating the
$yz$-component of the Reynolds stress against the corresponding component of
the mean-field strain tensor.
The current setup differs from that in \cite{KS22} in that a large-scale velocity is not imposed, and therefore, no off-diagonal Reynolds stress
is present.
The emergence of such off-diagonal components in shear flows was studied by \cite{Mitra+09}, who found 
an increase of the turbulent magnetic diffusivity.

\subsection{Simulations, data, and error bars}

We use the \textsc{Pencil Code} for our simulations \citep{JOSS}.
It uses sixth-order accurate spatial derivatives and a third-order time-stepping scheme.
It also allows us to compute turbulent transport coefficients with the test-field method.
For that purpose, we invoke the modules \texttt{testfield\_z} and \texttt{testscalar} within the \textsc{Pencil Code}.

We present our results for $\alpha$, $\etat$, and $\kappat$ in normalized form
and divide $\alpha$ by $A_0=\urms/3$ and $\etat$ and $\kappat$ by $D_0=\urms/3\kf$.
This allows us to compare runs with different rms velocity amplitudes.

Our results for the turbulent transport coefficients are functions of $z$ and $t$.
Since the turbulence in our simulations is homogeneous, we average the
resulting transport coefficients over $z$.
The resulting time series is then averaged over statistically steady intervals,
and error bars have been estimated by taking the largest departure to the average
from any one-third of the full time series.
For sufficiently long time series, the resulting errors are rather small,
so we often exaggerate them by a factor of 3 or 4.

\section{Results}

\subsection{Passive scalar results and comparison}

Although the results for $\etat$ have already been computed in
\cite{BSR17}, we compute them here again by invoking similar test-field
routines in the \textsc{Pencil Code} at the same time.
The test-field method for passive scalars was already described in
\cite{BSV09}.
\cite{R+11} applied it to passive scalar diffusion in compressible flows.
In the following, we use $\urms$ and $\kf$ to express our results in
nondimensional form by normalizing the diffusivities by $\urms/3\kf$.
Using earlier test-field results, this was found to be an accurate
estimate \citep{Sur+08}.

\begin{figure}[t]\begin{center}
\includegraphics[width=\columnwidth]{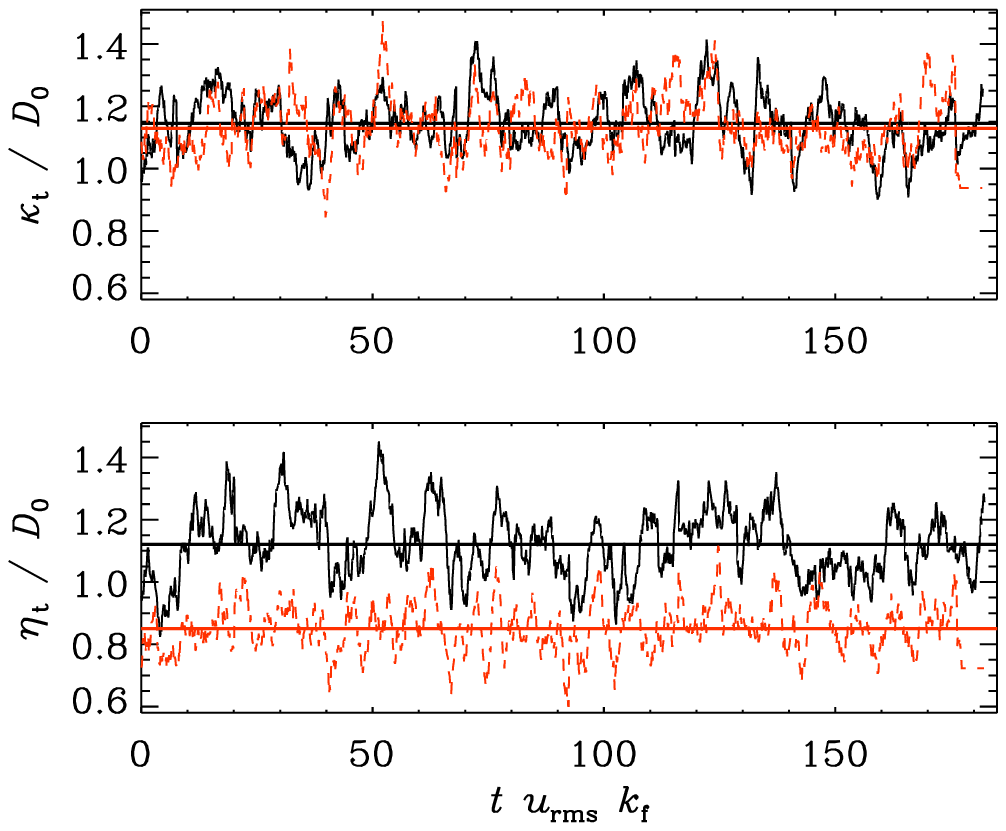}
\end{center}\caption{
Time series of $\kappat$ (upper panel) and 
$\etat$ (lower panel) for Runs~A without helicity (solid black line)
and with helicity (dashed red line) with $\Rey=2.4$.
The thick black and red horizontal lines denote the time-averaged values.
}
\label{pcomp}
\end{figure}

\begin{figure}[t]\begin{center}
\includegraphics[width=\columnwidth]{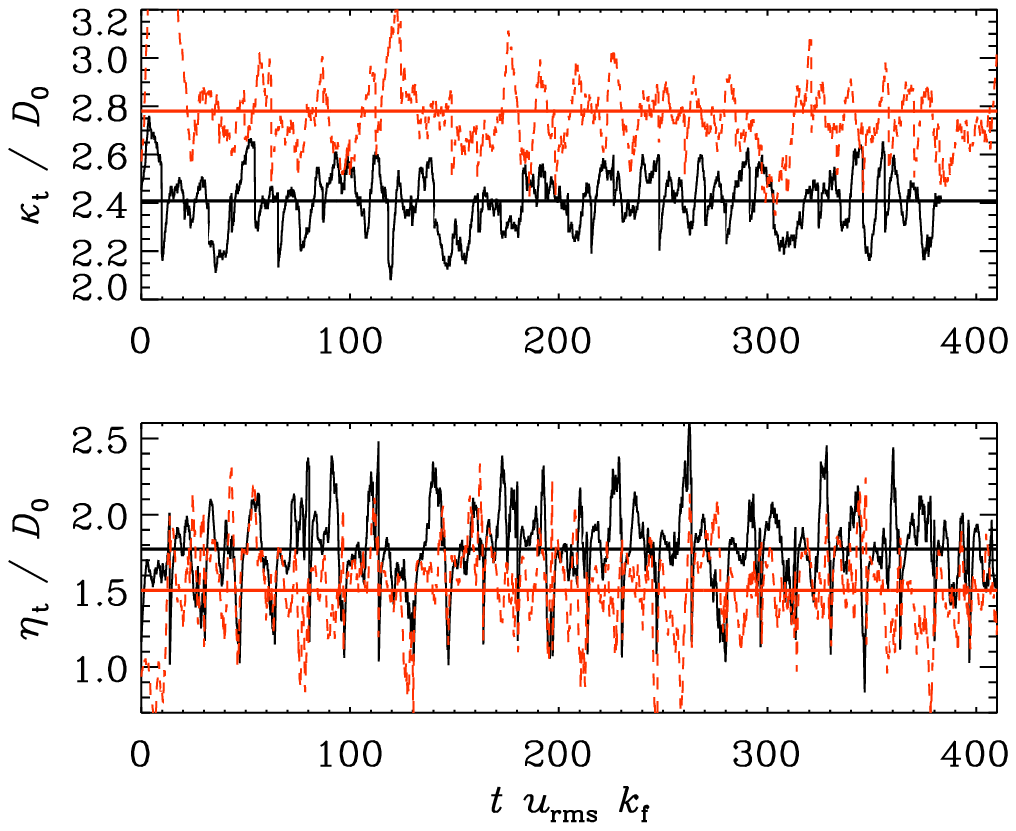}
\end{center}\caption{
Similar to \Fig{pcomp}, but for Runs~F with $\Rey=120$.
}\label{pcomp2}\end{figure}

\begin{figure}[t]\begin{center}
\includegraphics[width=\columnwidth]{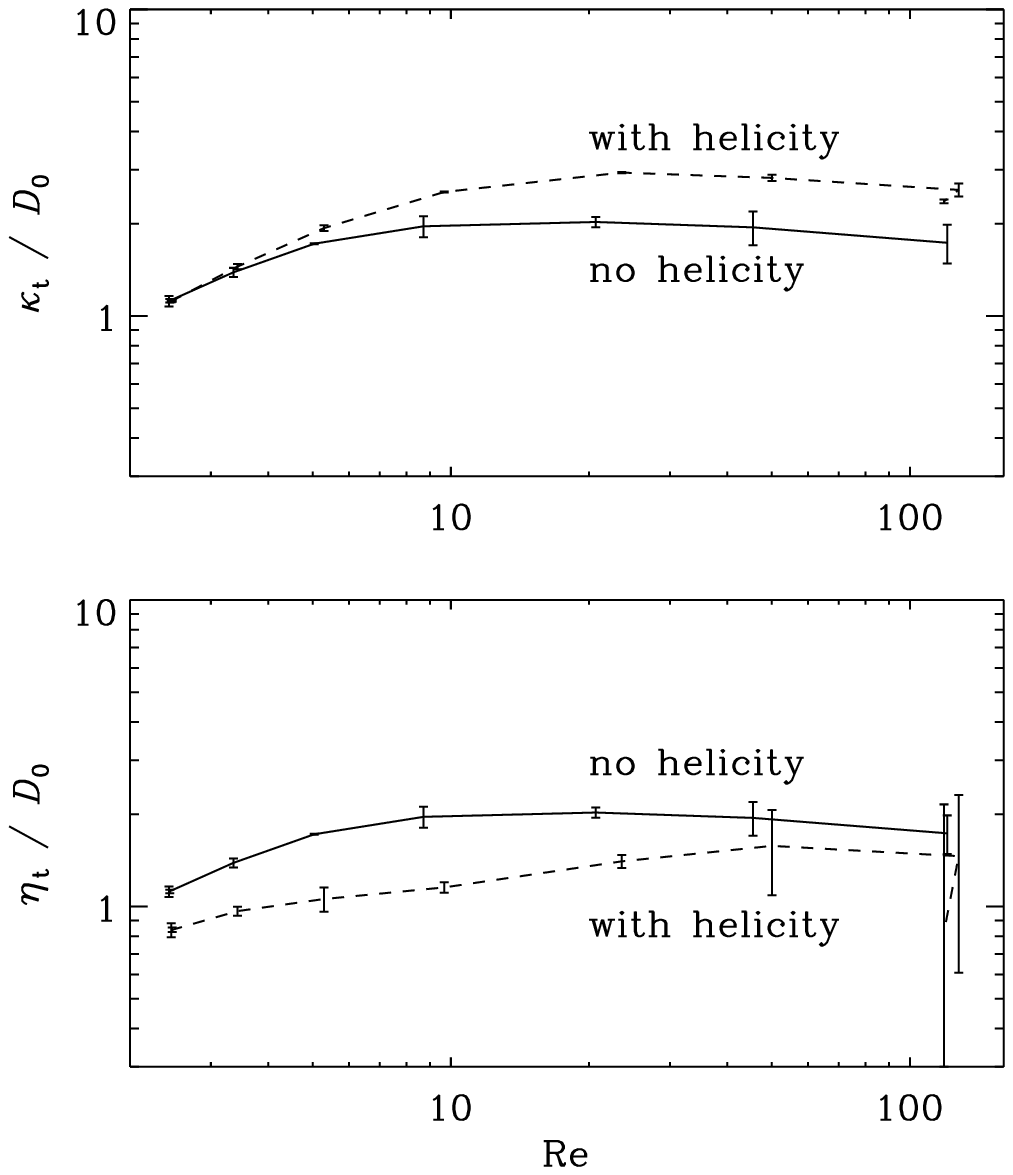}
\end{center}\caption{
Reynolds number dependence of $\kappat$ (upper panel) and 
$\etat$ (lower panel) for nonhelical (solid lines)
and helical (dashed lines).
The error bars have been exaggerated by a factor of 3.
}\label{presults}\end{figure}

\Fig{pcomp} shows a comparison of time series of $\kappat$ and $\etat$
for nonhelical and helical cases and $\Rey=2.4$.
While $\kappat$ appears to be unaffected by the presence of helicity,
$\etat$ is suppressed, as already found by \cite{BSR17}.
For $\Rey=120$, however, $\kappat$ is found to be \textit{enhanced}
by the presence of helicity; see \Fig{pcomp2}.
We have considered a number of additional simulations with other values of $\Rey$.
The dependence on $\Rey$ is shown in \Fig{presults};
see also \Tab{TSummary} for a summary.
The trend in $\etat$ does not follow a smooth dependence, suggesting that statistical noise or other unaccounted-for factors may have influenced the results.

The forcing is kept constant between different runs, so the resulting
rms velocity depends on how stiff the system is against this forcing.
We see that the value of the Mach number increases slightly with
increasing values of the Reynolds number.
We also see that the Mach number is slightly enhanced in the simulations with helical forcing.
This suggests that such flows are less effective in dissipating energy.
These slight changes in $\Ma$ do not significantly affect our results for $\etat$ and $\kappat$,
because we always present our results in normalized form and we
are here only interested in subsonic turbulence.
Note also that the compressibility of the turbulence affects only nonhelical contributions to the turbulent diffusion
\citep{RKB18}.

In \App{Comparison}, we compare our results with different degrees of
helicity with earlier simulations of rotating stratified turbulence in
which helicity is automatically being produced in a self-consistent way.
It turns out, however, that the enhancement of turbulent diffusion by
helicity is not being reproduced in such simulations.
We argue that this is caused by the more dominant effect of rotation
that strongly suppresses turbulent transport.

\subsection{Active scalar results}

\begin{figure}[t]\begin{center}
\includegraphics[width=\columnwidth]{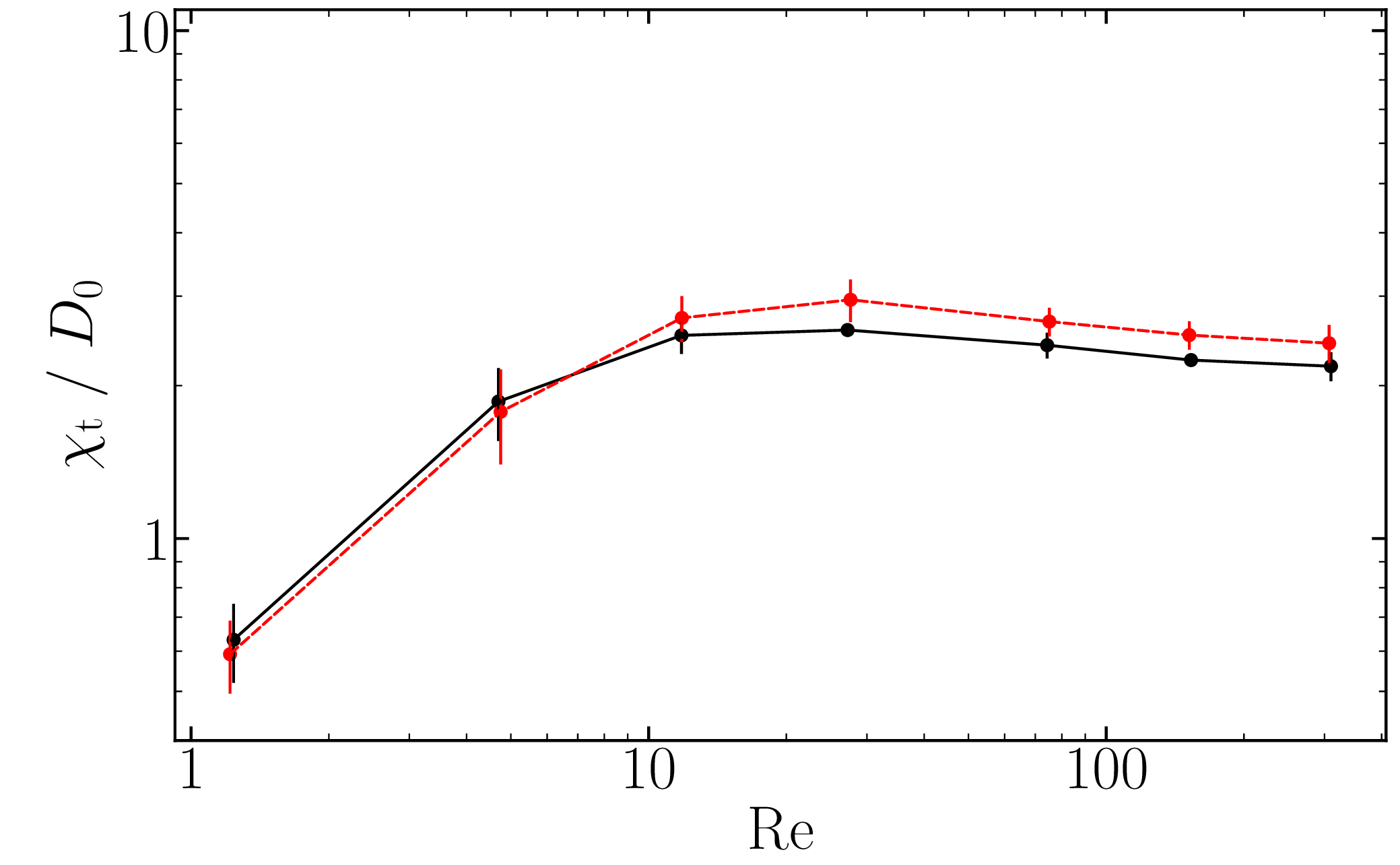}
\end{center}\caption{
Dependence of $\chit$ for nonhelical (black symbols, solid line) and
fully helical turbulence (red symbols, dashed line) as a function of
Reynolds number $\Rey=\Pe/\Pr$ with $\Pr=1$ in all cases.
To make the error bars more visible, they have been exaggerated by a factor of 4.
}\label{plot_chit_hel}\end{figure}

The results of simulations similar to those of \cite{KS22} are shown in \Fig{plot_chit_hel} and \Tab{cSummary}.
Here we see the turbulent heat diffusivity 
computed from an imposed entropy gradient \citep[see][for details]{KS22}
for nonhelical and helical cases.
For $\Pe = \Rey > 10$, there is a statistically significant increase of
$\chit$ by about 10\% for the helical cases
relative to the nonhelical ones.
These results were obtained by correlating the actual enthalpy flux
with the mean-field expression given by \Eq{equ:Fenth}. In \cite{KS22},
an alternative independent method was used where the mean entropy profile is
initially forced and then allowed to decay.
This yielded very similar results.

\begin{table*}\caption{
Values of $\kappat^\mathrm{nhel}$ and $\kappat^\mathrm{hel}$
as well as $\etat^\mathrm{nhel}$ and $\etat^\mathrm{hel}$,
normalized by $D_0\equiv\urms/3\kf$,
for the nonhelical and helical cases,
and $\alpha^\mathrm{hel}$ normalized by $A_0\equiv\urms/3$,
for the helical cases for different values of $\Rey$.
The value of $\Ma$ is given for completeness.
}\vspace{8pt}\begin{center}\begin{tabular}{lcccccccccc}
Run & $\Rey$ & $\kappat^\mathrm{nhel}/D_0$ & $\kappat^\mathrm{hel}/D_0$ & $\etat^\mathrm{nhel}/D_0$ & $\etat^\mathrm{hel}/D_0$ & $\alpha^\mathrm{hel}/A_0$ & $\Ma^\mathrm{nhel}$ & $\Ma^\mathrm{hel}$ \\
\hline
A & $  2.4$ & $   1.14\pm  0.01$ & $   1.12\pm  0.01$ & $   1.12\pm  0.01$ & $  0.84 \pm 0.015$ & $  -0.83\pm  0.01$ & $0.062$ & $0.063$ \\
B & $  3.4$ & $   1.47\pm  0.03$ & $   1.45\pm  0.03$ & $   1.39\pm  0.02$ & $  0.97 \pm 0.011$ & $  -0.95\pm  0.02$ & $0.068$ & $0.070$ \\
C & $  5.0$ & $   1.87\pm  0.04$ & $   1.93\pm  0.04$ & $   1.72\pm  0.00$ & $   1.06\pm  0.03$ & $  -1.02\pm  0.02$ & $0.077$ & $0.081$ \\
D & $  8.7$ & $   2.26\pm  0.03$ & $   2.54\pm  0.01$ & $   1.96\pm  0.05$ & $   1.15\pm  0.02$ & $  -0.96\pm  0.01$ & $0.089$ & $0.099$ \\
E & $ 20.7$ & $   2.54\pm  0.02$ & $   2.94\pm  0.01$ & $   2.03\pm  0.03$ & $   1.40\pm  0.02$ & $  -0.83\pm  0.02$ & $0.105$ & $0.120$ \\
F & $ 45.5$ & $   2.50\pm  0.03$ & $   2.82\pm  0.07$ & $   1.95\pm  0.08$ & $   1.58\pm  0.16$ & $  -0.75\pm  0.03$ & $0.116$ & $0.128$ \\
G & $120.6$ & $   2.27\pm  0.01$ & $   2.58\pm  0.12$ & $   1.73\pm  0.08$ & $   1.46\pm  0.28$ & $  -0.69\pm  0.07$ & $0.123$ & $0.130$ \\
\label{TSummary}\end{tabular}\end{center}\end{table*}

\begin{table}\caption{
Values of $\chit^\mathrm{nhel}$ and $\chit^\mathrm{hel}$
normalized by $D_0\equiv\urms/3\kf$,
for the nonhelical and helical cases.
The value of $\Ma$ is given for completeness.
}\vspace{8pt}\begin{center}\begin{tabular}{lcccc}
Run & $\Rey$ & $\chit^\mathrm{nhel}/D_0$ & $\chit^\mathrm{hel}/D_0$ & $\Ma$ \\
\hline
A &   $1.2$ & $0.63 \pm 0.03$ & $0.59 \pm 0.02$ & $0.031$ \\
B &   $4.7$ & $1.86 \pm 0.08$ & $1.78 \pm 0.09$ & $0.048$ \\
C &  $11.8$ & $2.51 \pm 0.05$ & $2.72 \pm 0.07$ & $0.060$ \\
D &  $27.6$ & $2.57 \pm 0.01$ & $2.95 \pm 0.07$ & $0.070$ \\
E &  $75.1$ & $2.40 \pm 0.03$ & $2.67 \pm 0.04$ & $0.077$ \\
F & $151.9$ & $2.25 \pm 0.02$ & $2.52 \pm 0.04$ & $0.077$ \\  
G & $307.0$ & $2.18 \pm 0.04$ & $2.42 \pm 0.05$ & $0.078$ \\
\label{cSummary}\end{tabular}\end{center}\end{table}

\begin{figure}[t]\begin{center}
\includegraphics[width=\columnwidth]{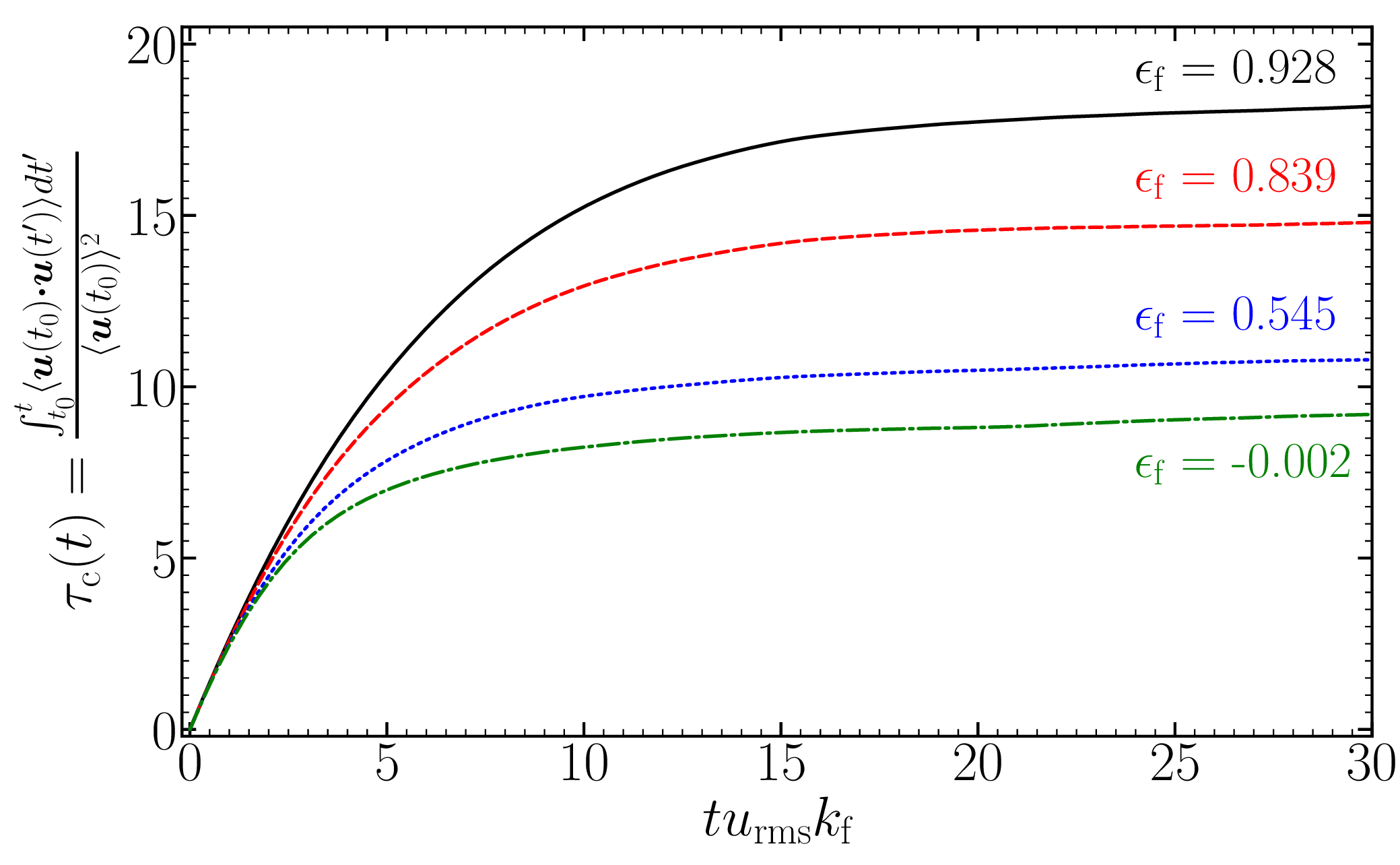}
\end{center}\caption{
Correlation time of turbulence computed from time integrals of velocity autocorrelation from runs with $\Rey=13$ and different
relative helicity $\epsilon_{\rm f}=
\langle\bm{u}\bm\cdot\bm\omega\rangle/\kf\urms^2$.
}
\label{plot_corr_hel}\end{figure}

The kinetic helicity effects on turbulent diffusion
of the mean magnetic and scalar fields are partially related to the helicity effect on the effective correlation time.
To examine this in more detail, we compute the correlation times as the late-time limit of
\begin{equation}
\tauc(t)=\left.\int_{t_0}^t\bbra{\uu(t_0)\cdot\uu(t')}\,\dd t'\right/\bbra{\uu^2(t_0)}.\label{equ:tauc_corr}
\end{equation}
The result is shown in \Fig{plot_corr_hel} for simulations with $\Rey=13$ and different values of the relative helicity.
We see that, through the presence of kinetic helicity,
the correlation time of the turbulent velocity field increases and is more than double as the kinetic helicity is increased from zero to one.
We note that the Reynolds number of these simulations is very modest.
Further studies at larger Reynolds numbers would be needed to establish the dependence of the correlation time on the kinetic helicity in more turbulent regimes; see \cite{RKB25}.

\begin{figure}[t]\begin{center}
\includegraphics[width=\columnwidth]{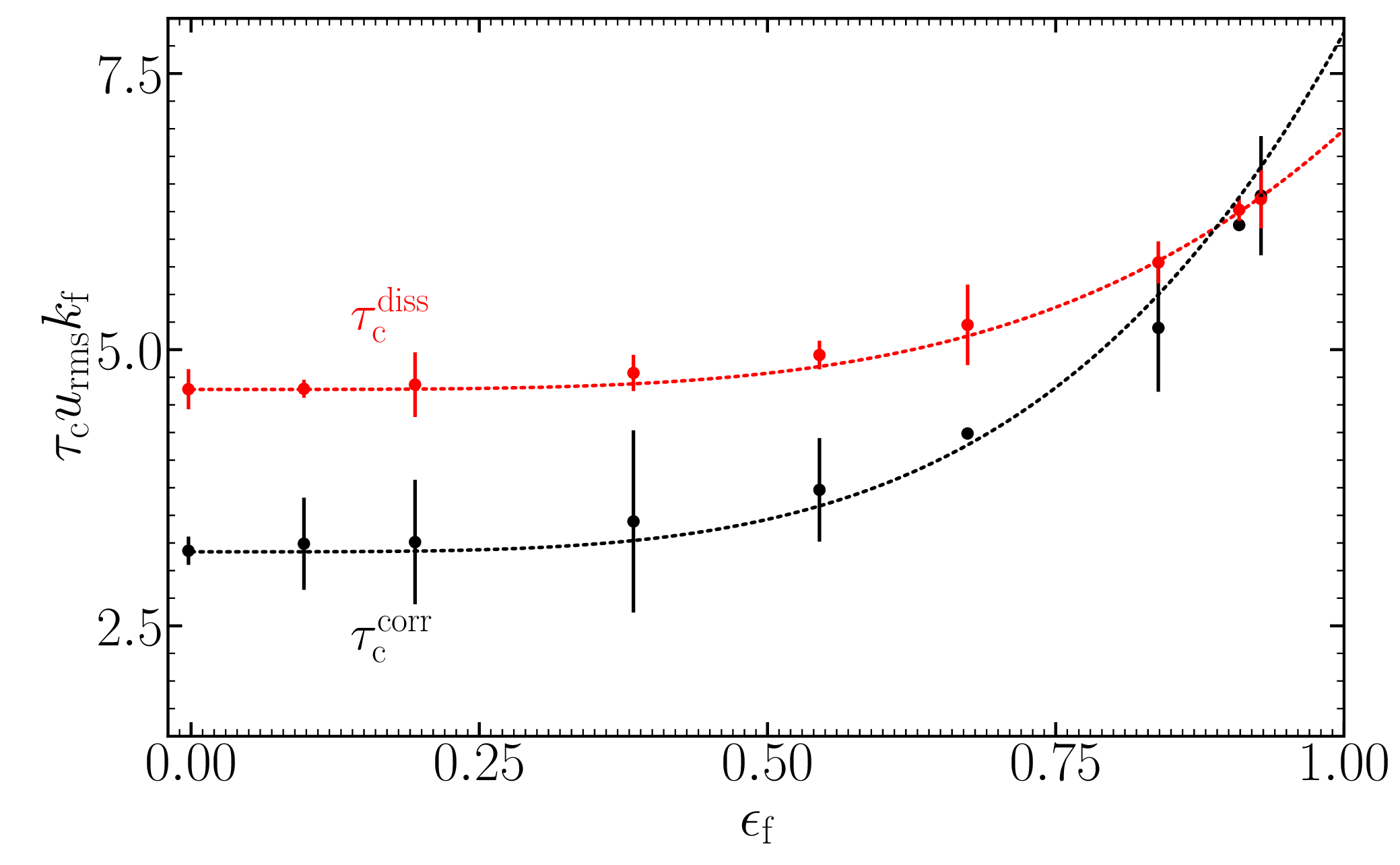}
\end{center}\caption{
Correlation time $\tauc$ as function of $\epsilon_{\rm f}$ from the late-time limit of \Eq{equ:tauc_corr} (black symbols) and from \Eq{equ:tauc_epsK} (red symbols) normalized by the turnover time
$(\urms\kf)^{-1}$ for the same runs as in \Fig{plot_corr_hel}.
The dotted lines are proportional to $\epsilon_{\rm f}^4$, and the error bars are boosted by a factor of ten for $\tauc^{\rm diss}$ and by five for $\tauc^{\rm corr}$.}
\label{plot_tauc}
\end{figure}

Another way to estimate the correlation time is obtained from the
ratio of turbulent kinetic energy and its dissipation rate:
\begin{eqnarray}
\tauc = \frac{\EK}{\epsK},\label{equ:tauc_epsK}
\end{eqnarray}
where $\EK= \onehalf \langle \uu^2 \rangle$ and $\epsK = 2\nu \langle
\SSSS^2 \rangle$. The results for the correlation time are summarized
in \Fig{plot_tauc}. Both measures of $\tauc$ show an increasing trend as a function of the fractional helicity, $\epsilon_{\rm f} = \overline{\bm{u}\bm\cdot\bm\omega}/\kf\urms^2$.

Regarding the usage of the energy dissipation rate $\epsilon_{\rm{K}}$ for the timescale arguments of turbulence, we should note the following points.
Although developed turbulence contains a very wide range of scales, it is still meaningful to use $\epsilon_\mathrm{K}$ for the arguments of turbulence timescale.
In the inertial range of fully developed turbulence, we have a local equilibrium between the production rate of turbulent energy and its dissipation rate.
In this range, the dissipation rate is equivalent both to the energy injection rate at the integral scale and to the energy flux (the spectral energy transfer from larger scale to smaller scale).
In this sense, the energy dissipation rate $\epsilon_{\rm{K}}$ is the most appropriate turbulence statistical quantity that describes the timescale of turbulence.
This is the reason why we also adopt $\epsilon_\mathrm{K}$ in the timescale argument of turbulence.

Turbulent transport coefficients depend on some statistical quantities such as the turbulent energy $E_\mathrm{K}$, its dissipation rate $\epsilon_{\rm{K}}$, 
kinetic helicity $H_{\rm{K}}$, etc, as well as the time and/or length scales of turbulence, which are determined by $E_\mathrm{K}$, $\epsilon_\mathrm{K}$ and $H_\mathrm{K}$,
as well as the velocity strain rate, vorticity, pressure, etc. 
Generally, the kinetic helicity $H_\mathrm{K}$ depends on the vorticity/rotation and density stratification or turbulence inhomogeneity as well as the external forcing.
Here, for simplicity, we assume that deviations of the turbulence timescale from the usual eddy turnover time,
$\tau_{\rm{c}} = E_{\rm{K}}/ \epsilon_{\rm{K}}$, can be expressed in terms of the kinetic helicity as $\tau_{\rm{c}}(H_{\rm{K}})$,
and examine the dependence of $\tau_\mathrm{c}$ on $H_\mathrm{K}$.

\subsection{Comparisons with the theoretical predictions}
\label{theory}

\begin{figure}
\centering
\includegraphics[width=8.0cm]{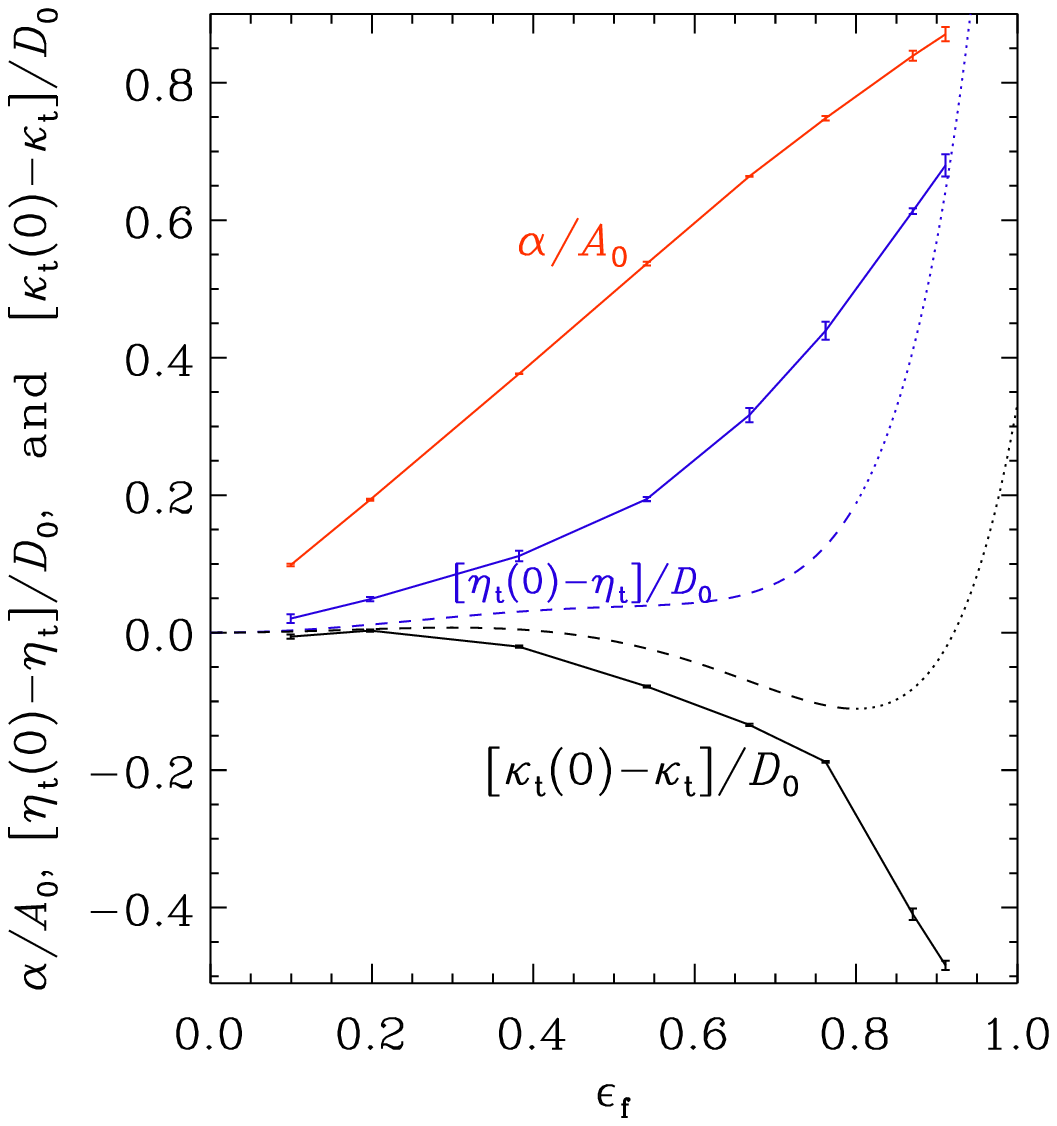}
\caption{\label{Fig7}
Dependencies of $\alpha$ (red solid line), 
$\eta_\mathrm{t}(0) -\eta_\mathrm{t}$ (blue solid line) and
$\kappa_\mathrm{t}(0) -\kappa_\mathrm{t}$ (black solid line)
on the fraction $\epsilon_{\rm f}$ of the kinetic helicity for $\Rey=13$.
The theoretical dependencies given by Equations~(\ref{MT1})--(\ref{scaling-tau_c}) are shown by dashed and dotted blue and black curves.
The theoretical results for $\epsilon_\mathrm{f}\ga0.8$ are shown as dotted lines, because they may not be reliable.
}
\end{figure}

Let us compare the obtained numerical results
with the theoretical predictions by \cite{RKB25}, 
where the path-integral approach for a random velocity field with a finite correlation time has been used.
According to the theory, the turbulent magnetic diffusion coefficient $\eta_{_{\rm t}}(H_{\rm K})$ is given by
\begin{eqnarray}
\eta_{_{\rm t}}(H_{\rm K}) =\eta_{_{\rm t0}} \, {\tau_{\rm c}(H_{\rm K}) \over \tau_0} \left(1 - {\tau_{\rm c}^2(H_{\rm K}) \over 3} \, {H_{\rm K}^2 \over \langle {\bm u}^2 \rangle}\right) ,
\label{MT1}
\end{eqnarray}
while the turbulent diffusion coefficient $\kappa_{_{\rm t}}(H_{\rm K})$ of the scalar field is 
\begin{eqnarray}
\kappa_{_{\rm t}}(H_{\rm K}) = \kappa_{_{\rm t0}} \, {\tau_{\rm c}(H_{\rm K}) \over \tau_0} \left(1 - {\tau_{\rm c}^2(H_{\rm K}) \over 6} \, {H_{\rm K}^2 \over \langle {\bm u}^2 \rangle}\right) ,
\label{ST1}
\end{eqnarray}
where $H_{\rm K}=\langle{\bm u} \cdot  {\bm \omega}\rangle$,  $\eta_{_{\rm t0}} = \eta_{_{\rm t}}(H_{\rm K}=0)$,
$\kappa_{_{\rm t0}} = \kappa_{_{\rm t}}(H_{\rm K}=0)$  and 
$\tau_0 =\tau_{\rm c}(H_{\rm K}=0)= (\urms\kf)^{-1}$.
Applying two independent methods (based on the noninstantaneous correlation functions and the rate of energy dissipation)
for the calculation of the correlation time versus the fraction of kinetic helicity, our numerical results suggest that 
\begin{eqnarray}
\tau_{\rm c}(H_{\rm K}) / \tau_0 \approx 1 + 0.5 \epsilon_{\rm f}^4 .
\label{scaling-tau_c}
\end{eqnarray}
Using Equations~(\ref{MT1})--(\ref{scaling-tau_c}), we plot in \Fig{Fig7} the dependences
$\eta_\mathrm{t}(0) -\eta_\mathrm{t}$ and $\kappa_\mathrm{t}(0) -\kappa_\mathrm{t}$ on the fraction $\epsilon_{\rm f}$ of the kinetic helicity for $\Rey=13$.
Here, $\eta_\mathrm{t}(0) =\eta_\mathrm{t}(\epsilon_{\rm f}=0)$ and
$\kappa_\mathrm{t}(0) =\kappa_\mathrm{t}(\epsilon_{\rm f}=0)$,
and $\alpha$ is normalized by $A_0=u_{\rm rms}/3$,
while turbulent diffusion coefficients are 
normalized by $D_0=u_{\rm rms}/3k_{\rm f}$,
where $k_{\rm f}$ is the forcing wavenumber.
The theoretical dependencies given by Equations~(\ref{MT1})--(\ref{scaling-tau_c}) are shown as dashed and dotted blue and black curves.
The theoretical results for $\epsilon_\mathrm{f}\ga0.8$ are shown as dotted lines, because they may not be reliable.

As follows from \Fig{Fig7}, the turbulent magnetic diffusion coefficient is reduced by the kinetic helicity,
while the turbulent diffusion coefficient for the scalar field is increased by the kinetic helicity.
These arguments can explain the results of our direct numerical simulations; see also Fig.~1 for $\Rey=120$ in \cite{RKB25}.

Using an approach based on the Furutsu--Novikov theorem \citep{Furutsu63, Novikov65}, \cite{KS25}
found that the turbulent diffusivities of both the mean passive scalar and the mean magnetic field are suppressed by the kinetic helicity.
Note that \cite{KS25} have not taken into account 
the dependence of the correlation time on the kinetic helicity.
This may explain the discrepancy with our numerical results
related to the helicity effect on the turbulent diffusion of the scalar fields.

\subsection{Scale dependence}
\label{ScaleDependence}

\begin{figure}[t]\begin{center}
\includegraphics[width=\columnwidth]{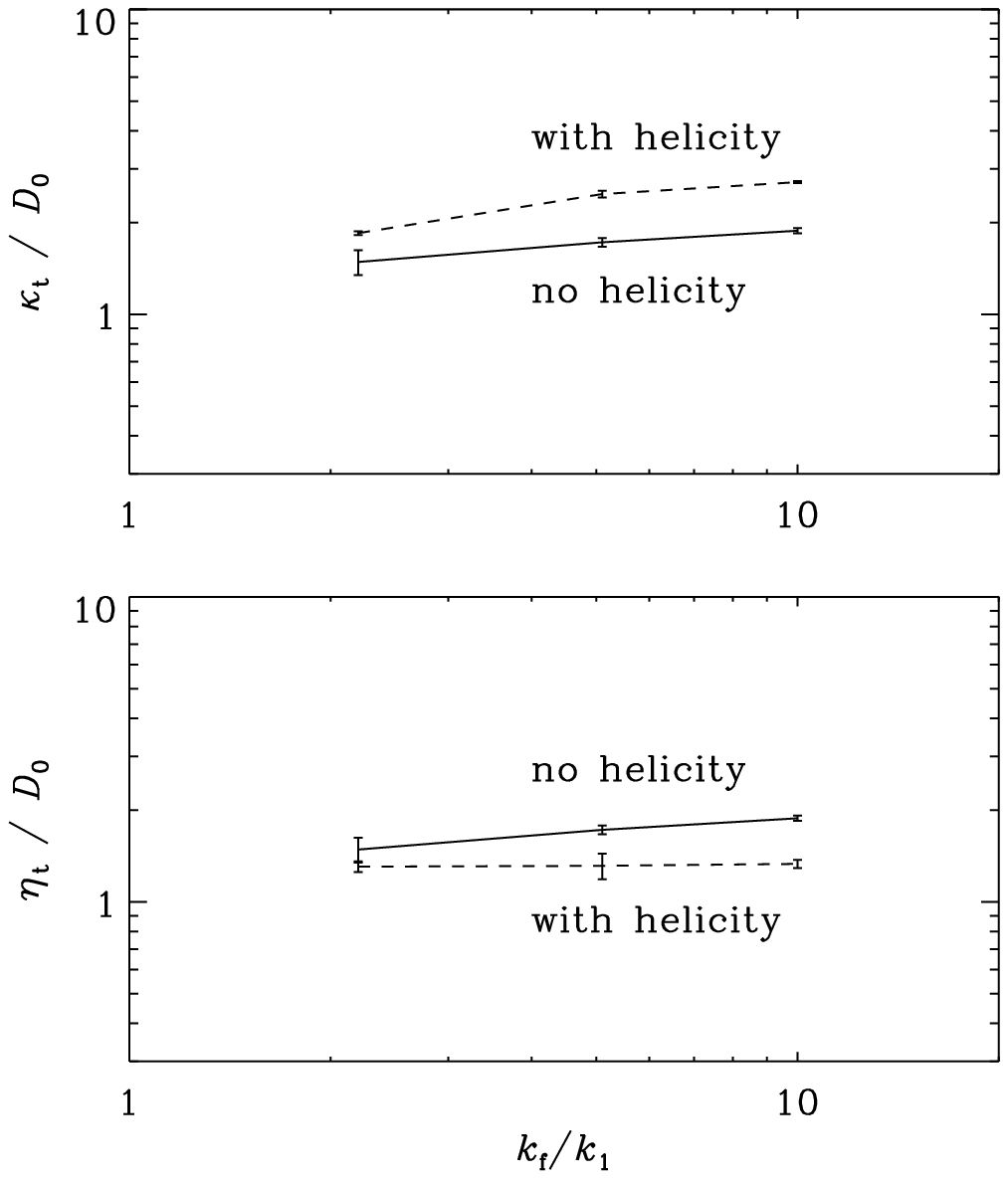}
\end{center}\caption{
Scale separation dependence of $\kappat$ (upper panel) and 
$\etat$ (lower panel) for nonhelical (solid lines)
and helical (dashed lines).
To make the error bars more visible, they have been exaggerated by a factor of 4.
}\label{presults_kf}\end{figure}

To assess the scale dependence of the difference of turbulent transport for helical and nonhelical cases, we have varied the ratio $\kf/k_1$, keeping the viscosity constant.
This implies that $\Rey$ decreases with increasing $\kf$.
In all cases, we used $512^3$ meshpoints.
The results for helical and nonhelical turbulence are compared in \Fig{presults_kf} and \Tab{TSummary_kf}.
We see that there is a slight increase in the difference between helical and nonhelical cases.
For $\kappat$, however, the difference between helical and nonhelical cases is rather weak.

\begin{table*}\caption{
Values of $\kappat^\mathrm{nhel}$ and $\kappat^\mathrm{hel}$
as well as $\etat^\mathrm{nhel}$ and $\etat^\mathrm{hel}$,
normalized by $D_0\equiv\urms/3\kf$,
for the nonhelical and helical cases,
and $\alpha^\mathrm{hel}$ normalized by $A_0\equiv\urms/3$,
for the helical cases for different values of $\kf$.
The value of $\Ma$ is given for completeness.
}\vspace{2pt}\begin{center}\begin{tabular}{lccccccccccc}
Run & $\kf/k_1$ & $\Rey$ & $\kappat^\mathrm{nhel}/D_0$ & $\kappat^\mathrm{hel}/D_0$ &
$\etat^\mathrm{nhel}/D_0$ & $\etat^\mathrm{hel}/D_0$ & $\alpha^\mathrm{hel}/A_0$ & $\Ma^\mathrm{nhel}$ & $\Ma^\mathrm{hel}$ \\
\hline
a & $  2.2$ & $281.1$ & $   1.66\pm  0.03$ & $   1.85\pm  0.03$ & $   1.49\pm  0.14$ & $   1.31\pm  0.05$ & $  -0.65\pm  0.01$ & $0.125$ & $0.131$ \\
b & $  5.1$ & $120.6$ & $   2.27\pm  0.01$ & $   2.48\pm  0.06$ & $   1.72\pm  0.06$ & $   1.31\pm  0.13$ & $  -0.69\pm  0.01$ & $0.123$ & $0.130$ \\
c & $ 10.0$ & $ 59.3$ & $   2.49\pm  0.01$ & $   2.71\pm  0.02$ & $   1.88\pm  0.04$ & $   1.33\pm  0.04$ & $  -0.76\pm  0.00$ & $0.119$ & $0.129$ \\
\label{TSummary_kf}\end{tabular}\end{center}\end{table*}

\section{Conclusions}

Our simulations have revealed a surprising difference in the helicity effect for passive and active scalars on the one hand and magnetic fields on the other.
As for magnetic fields, the helicity effect on the turbulent diffusivity does not exist for small Reynolds numbers.
Above Reynolds numbers of about 20, it does not change much anymore, and there is no indication that it disappears at larger values.

The key numerical result of the present study is the enhancement of turbulent diffusion of the mean passive and active scalar fields by the kinetic helicity.
This result is opposite to the magnetic case where turbulent magnetic diffusion is decreased by the kinetic helicity.
We also found that the correlation time of the turbulent velocity field increases because of kinetic helicity.
The latter is one of the main points relevant for understanding the kinetic helicity effects on turbulent diffusion of scalar and magnetic fields (see Section~\ref{theory}).

The enhancement of the turbulent passive scalar diffusion examined here can be compared with the effect of rotation and stratification on the passive scalar diffusivity.
As discussed in the introduction, rotating stratified turbulent flows also attain kinetic helicity and for such flows, it was previously found that the passive scalar diffusivity gets reduced as the rotation speed is
increased, just like the magnetic diffusivity, which also became smaller
\citep{BRK12}.
This effect was not ascribed to the presence of helicity, but it was simply regarded as a rotational suppression of
the magnetic diffusivity.
This difference can probably be explained by the anisotropy of the flow that is being produced in rotating stratified turbulence, which is a more
complicated situation than just a helically forced flow.

Qualitatively, one could understand the helicity effect on the magnetic field as a tendency to support dynamo action, or, alternatively, as a tendency for rotational suppression of the magnetic diffusivity.
For passive and active scalars, on the other hand, there is no dynamo effect.
Furthermore, in some special deterministic flows
(the Roberts-IV flow; see \cite{DBM13}), the effective magnetic diffusivity can
even be negative and thereby lead to dynamo action.
Such an effect was never found for passive or active 
scalars or magnetic fields in turbulent flows at high Reynolds numbers.
What has been previously found, 
however, is a suppression of both $\etat$ and $\kappat$ for potential (compressible) flows \citep{R+11,RKB18}.
In the present work, however, we have only considered nearly incompressible flows for actual turbulence simulations, as opposed to
some constructed flows such as the Roberts flow.

\begin{acknowledgements}
We thank Nathan Kleeorin for useful discussions and the referees for their reports.
We also acknowledge the discussions with participants
of the Nordita Scientific Program on ''Stellar Convection: Modelling, Theory and Observations", Stockholm (September 2024).
This research was supported in part by the
Swedish Research Council (Vetenskapsr{\aa}det) under Grant No.\ 2019-04234,
the National Science Foundation under grant no.\ NSF AST-2307698,
a NASA ATP Award 80NSSC22K0825, and
the Munich Institute for Astro-, Particle and BioPhysics (MIAPbP), which
is funded by the Deutsche Forschungsgemeinschaft (DFG, German Research
Foundation) under Germany´s Excellence Strategy - EXC-2094 - 390783311.
Part of this work was supported by the Japan Society of the Promotion of Science (JSPS) Grants-in-Aid for Scientific Research JP23K25895.
We acknowledge the allocation of computing resources provided by the
Swedish National Allocations Committee at the Center for
Parallel Computers at the Royal Institute of Technology in Stockholm, and by the North German Supercomputing Alliance (HLRN) in G\"ottingen and Berlin, Germany.

\vspace{2mm}\noindent
{\em Software and Data Availability.} The source code used for
the simulations of this study, the {\sc Pencil Code} \citep{JOSS},
is available on \url{https://github.com/pencil-code/}.
The simulation setups and corresponding secondary data are available on
\dataset[http://doi.org/10.5281/zenodo.15083000]{http://doi.org/10.5281/zenodo.15083000}.
\end{acknowledgements}

\appendix
\section{Comparison with earlier work}
\label{Comparison}

\begin{figure}[t]\begin{center}
\includegraphics[width=\columnwidth]{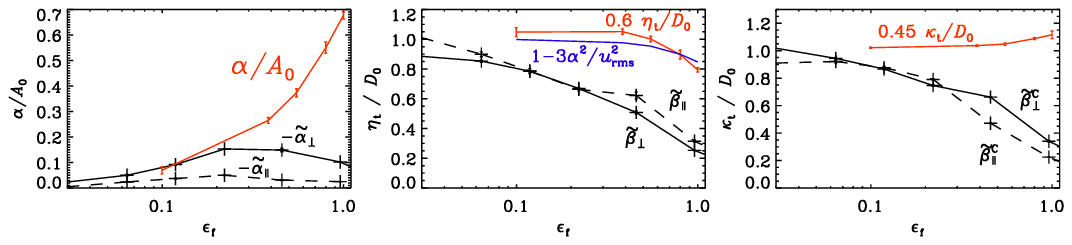}
\end{center}\caption{
Dependence of the fractional helicity, $\epsf$, and comparison of
the values of $\alpha$ and the turbulent magnetic and passive scalar
diffusivities with the earlier work of \cite{BRK12}, in which kinetic
helicity is being produced by the interaction with rotation and
stratification.
The originally used symbols of \cite{BRK12} have been retained:
$-\tilde{\alpha}_\perp$ and $-\tilde{\alpha}_\|$ for the normalized
perpendicular and parallel components of the $\alpha$ effect,
$\tilde{\beta}_\perp$ and $\tilde{\beta}_\|$ for those of the
magnetic diffusivity, and $\tilde{\beta}_\perp^\mathrm{C}$ and
$\tilde{\beta}_\|^\mathrm{C}$ for those of the passive scalar diffusivity.
The tildes indicate appropriate normalization.
In the second panel, we also show in blue $1-3\alpha^2/\urms^2$.
}\label{psummary_Odep_g08_compare}\end{figure}

In \Fig{psummary_Odep_g08_compare}, we compare the values of
the $\alpha$ effect and the turbulent magnetic and passive scalar
diffusivities with those of the earlier work of \cite{BRK12}, in which kinetic
helicity is being produced by the interaction with rotation and
stratification.
Here, we have estimated the fractional helicity from the product
of Coriolis number $\Co=2\Omega/\urms\kf$ and gravity number
$\Gr=1/H_\rho\kf$, where $\Omega$ is the angular velocity,
$\kf$ is the forcing wavenumber of the turbulence, and $H_\rho$
is the density scale height.
We used a formula by \cite{Jabb+14}, $\epsf=2\,\Co\,\Gr$.
For the present simulations, we used $\epsf\approx2\sigma/(1+\sigma^2)$.

There is not much agreement with our present simulations, shown in red.
This shows that other effects such as the rotational suppression of
turbulent transport play a more dominant role than just the helicity.

\bibliography{ref}{}
\bibliographystyle{aasjournal}
\end{document}